\documentclass[]{gSSR2e}

\begin{document}
\doi{10.1080/1744250YYxxxxxxxx}
 \issn{1744-2516}
\issnp{1744-2508}
\jvol{00} \jnum{00} \jyear{2009} \jmonth{January}

\markboth{L.~A. Bordag}{Stochastics: An International Journal}


\title{{\itshape Study of the risk-adjusted pricing methodology model with methods of Geometrical
Analysis}}

\author{L. A. Bordag$^{\ast}$\thanks{$^\ast$Corresponding author. Email:
Ljudmila.Bordag@hh.se \vspace{6pt}}\\\vspace{6pt}{\em{Halmstad
University, Box 823, 301 18 Halmstad,
Sweden}}\\\vspace{6pt}\received{October 2009} }

\maketitle

\begin{abstract}

\noindent {\bf Abstract.} Families of exact solutions are found to
a nonlinear modification of the Black-Scholes equation. This
risk-adjusted pricing methodology model (RAPM) incorporates both
transaction costs and the risk from a volatile portfolio. Using
the Lie group analysis we obtain the Lie algebra admitted by the
RAPM equation. It gives us the possibility to describe an optimal
system of subalgebras and correspondingly the set of invariant
solutions to the model. In this way we can describe the complete
set of possible reductions of the nonlinear RAPM model. Reductions
are given in the form of different second order ordinary
differential equations. In all cases we provide solutions to these
equations in an exact or parametric form. We discuss the
properties of these reductions and the corresponding invariant
solutions.

\begin{classcode} 35K55, 34A05, 22E60 \end{classcode}\bigskip

\end{abstract}

\section{Introduction}

One of the most important problems at present is  how to
incorporate both the transaction costs and the risk from a
volatile (unprotected) portfolio into the governing Black-Sholes
equation. In the pioneering work of Leland \cite{Leland1985},
devoted to the problem of option pricing in the presence of
transaction costs, the idea of a periodic revision of a hedging
portfolio was introduced. Leland assumed that the level of
transaction costs is a constant, i.e. we have a market with
proportional transaction costs. He reduced this problem to a
nonlinear partial differential equation with an adjusted
volatility. Leland claimed that the terminal value of the
portfolio approximates the payoff as the length of a revision
interval tends to zero. Later, Kabanov and Safarian
\cite{Kabanov1997}  proved that Leland's conjecture based on
approximate replication fails and his model has a non-trivial
limiting hedging error relative to simulated marked prices (see as
well the detailed discussion in \cite{Kabanov2010}). Mathematical
problems arise in the limiting cases as revisions become
unboundedly frequent. As a practical matter, extremely frequent
revisions will not be desirable and the average errors are less
than one-half of one per cent of the price suggested by Leland's
formula \cite{Leland2007}. Within this model Kratka \cite{Kratka}
has suggested a mathematical method for pricing derivative
securities in the presence of proportional transaction costs and
he additionally took into account the risk of the unprotected
portfolio in between the revisions. Janda{\v{c}}ka and
{{\v{S}}ev{\v{c}}ovi{\v{ c}}} \cite{JandarckaSevcovic} modified
Kratka's approach in order to derive a scale-invariant model.

In the model the risk from the volatile portfolio is described by
the average value of the variance of the synthesized portfolio.
The mathematical model was referred to as the risk-adjusted
pricing methodology (RAPM) model. The RAPM model  generalizes the
famous Black-Scholes model for pricing of derivative securities.
In the model setting both the transaction costs  and the
unprotected portfolio risk depend on the time interval between two
transactions and minimizing of the total risk leads to the RAPM
model. The model was studied recently with numerical methods in
the case of European and
American options \cite{Sevcovic}.\\
We describe briefly the model settings.

The authors assume that the stock price dynamics is given by the
geometric Brownian motion
\begin{equation} \label{browMot}
S_t=S_0 \exp{\left( (\rho -\sigma^2/2)t + \sigma W_t \right)},
\end{equation}
where $\{W_t, t \ge 0 \}$ is the Wiener process, $\rho \in
{\mathbb R}$ is the drift and $\sigma>0$ is the instantaneous
volatility of the asset, $\rho, \sigma$ are constants.
It is assumed that the risk-free bond earns at a continuously compounded constant rate $r$.\\
The time-steps $\Delta t $ at which the portfolio can be hedged
against the price change of the underlying asset $S_t$ are
non-infinitesimal  and fixed. Additionally, the authors introduce
the idea of a switching time $t^*$ for the last revision of the
portfolio. This means that the time interval $(0,T)$ is divided in
two parts, in the first part $(0,t^*)$ the revisions of portfolio
will be done regularly, and in the second one $(t^*, T)$ there are
no revisions and correspondingly no transaction costs. It is
assumed that the interval $(t^*,T)$ is very small and in this
interval the price of the contingent claim $ u(S,t),~~ t \in
[t^*,T]$ is defined as in the classical Black-Scholes formula
(here $T$ is the maturity time). It is assumed that the model
(similar to Leland's model)
does not include the cost of establishing the initial investor's portfolio composition.\\
At time  $t$ the value of the dynamically hedged portfolio $V_t$
is $V_t^{\phi} = \delta_t S_t + \beta_t B_t$, where $\delta_t$ is
a number of units of the stock (a constant on each time interval
$\Delta t$), $B_t$ is the value of the bond and $\beta_t$ is a
number of units of the bond. We can put $B_0=1$ without loss of
generality and rewrite the previous relation in the form
$V_t^{\phi} = \delta_t S_t + \beta_t e^{r t}$. The pair
$\phi=(\delta_t,\beta_t)$ defines the self-financing hedging
strategy that maintains the portfolio.

The change of $V_t^{\phi}$ in any time-step $\Delta t $ is equal
to $\Delta V_t^{\phi}= V_{t+\Delta t}^{\phi} - V_t^{\phi}= \beta_t
e^{r t}(e^{r \Delta t} -1)+ \delta_t (S_{t+\Delta t}-S_t) - r_R
S_{t} \Delta t$. The total risk premium $r_R $ contains two parts
$r_R=r_{TC}+r_{VP}$. The transaction costs (TC) in this case are
modeled by the expression
\begin{equation}r_{TC}= \frac{C \sigma S |u_{SS}|}{\sqrt{2
\pi} }, ~~ C=\frac{S_{ask}-S_{bid}}{S},
\end{equation}
where $C$ is the round trip transaction costs per unit dollar of
transaction \cite{Leland1985}, \cite{Hoggardt}, \cite{Kwok} and
$u(S,t)$ is the value function of the contingent claim with
respect to the asset price $S$ and time $t$. During the time-step
$\Delta t$ the portfolio is unprotected and the risk connected
with a volatile portfolio (VP) is modeled by
\begin{equation}
r_{VP}= \frac{1}{2}R \sigma^4 S^2 (u_{SS})^2 \Delta t,
\end{equation}
where $R$ is a risk premium coefficient introduced in
\cite{Kratka} and \cite{JandarckaSevcovic} and represents the
marginal value of investor's exposure to a risk. The total risk
premium depends on the time-lag $\Delta t$ and it is a strong
convex function between two consecutive portfolio revisions
\cite{Sevcovic}. To obtain a risk-adjusted Black-Scholes equation
the authors minimize the total risk premium $r_R=r_{TC}+r_{VP}$.
They then obtain for the optimal time-lag the following value
$$\Delta t_{opt}=
\frac{C^{2/3}}{\sigma^2(R \sqrt{ 2 \pi} |S u_{SS}|)^{2/3}}.$$
Using Ito's formula the authors of \cite{JandarckaSevcovic}
finally obtain the risk-adjusted pricing methodology model
\begin{equation} \label{sevcov}
u_t+\frac{1}{2}\sigma^2 S^2 u_{SS}(1-\mu (S
u_{SS})^{\frac{1}{3}})-r u + r S u_S=0,~~\mu= 3 \left(\frac{C^2
R}{2 \pi} \right)^{1/3},
\end{equation}
where $t \in (0,t^*)$ and the value $t^*$ is determined by the
implicit equation $T-t^*= \min_{S>0}{\Delta t_{opt}(S,t^*)}$.  The
equation represents a well-posed parabolic problem under the
condition that
\begin{equation} \label{parab}
S u_{SS}(S,t)< \left( \frac{3}{ 4 \mu } \right)^3.
\end{equation}
The condition (\ref{parab}) will not be fulfilled for usual Call
and Put options at $S=E$ and $t \to T^-$, where $E$ is the strike
price of the corresponding option. To avoid the singularities in
the model the authors introduced the switching time $t^*$ such
that condition (\ref{parab}) is satisfied by $t=t^*$.
The equation  for $t^*$ which can be reduced to the form $T-t^*=C
R^{-1}\sigma^{-2}$ (for European Call and Put options) has a
positive solution and the condition (\ref{parab}) is satisfied if
\begin{equation}
\frac{C}{R}< \sigma^2 T, ~~ C R <\frac{\pi}{8}.
\end{equation}
From the analytical point of view this model is represented by a
fully nonlinear parabolic differential equation (PDE) with a
singular perturbation. Our goal is the study of the RAPM model
with the methods of Geometrical Analysis.

\section{Symmetry properties}

Equation (\ref{sevcov}) is the main subject of our investigations.
The equation possesses a complicated analytical and algebraic
structure. We provide the Lie group analysis of this equation with
the goal of describing the complete set of symmetries of equation
(\ref{sevcov}) and to obtain possible reductions. Using the
invariants of the subgroups of the symmetry group of the studied
equation we reduce the partial differential equation to ordinary
differential equations (ODEs). Solutions to these ODEs give us the
invariant solutions to the nonlinear RAPM model in an
analytical form.\\
We obtain the symmetry group of the RAPM model in the way
suggested by Sophus Lie and developed further in
\cite{Ovsiannikov}, \cite{Olver} and \cite{Ibragimov}. We first
find, using the Lie determining equations, the Lie algebra $L_r$
of a dimension $r$ admitted by the equation. Then we use an
exponential map $\exp: L_r \to G_r$ and obtain the transformations
of the symmetry group $G_r$. To each subalgebra $h_i \subset L_r$
corresponds a subgroup $H_i$ of $G_r$ \cite{Ibragimov},
\cite{Olver}, \cite{Ovsiannikov}. In most cases we do not need the
explicit form of the group transformations and use directly the
subalgebras $h_i$ of $L_r$ in order to reduce the RAPM model.

In this way we  prove the following theorem.
{\begin{theorem} The
equation (\ref{sevcov}) admits a four dimensional Lie algebra
$L_4$ with the following infinitesimal generators
\begin{equation}
U_1=S\frac{\partial}{\partial S}+u \frac{\partial}{\partial u}, ~~
U_2=e^{r t}\frac{\partial}{\partial u},~~
U_3=\frac{\partial}{\partial t},~~ U_4=S \frac{\partial}{\partial
u}. \label{generatorsSev}
\end{equation}
The commutator relations are
\begin{eqnarray}
[U_1,U_2]=-U_2,~~~~~~~~~~~~~[U_2,U_3]=-r U_2,\label{commutator} \\
~[U_1,U_3]=[U_1,U_4]=[U_2,U_4]=[U_3,U_4]=0.
\end{eqnarray}
\end{theorem}
 The commutator relations (\ref{commutator}) depend
on the parameter $r$, i.e. on the interest rate included in the
model. Depending on whether $r=0$ or $r\ne 0$, we obtain different
commutation relations for the algebra generators of the Lie
algebra $L_4$. After the proper choice of generators
we obtain, in both cases, isomorphic algebras.\\
All four-dimensional real Lie algebras were classified by Patera
and Winternitzs  \cite{PateraWinternitzs1977}. We will use this
classification and the corresponding notations for generators of
$L_4$. The algebra is spanned by the following generators
$L_4=<~e_1,e_2,e_3,e_4~>$, which will have different meaning
depending on the value of $r$. We denote a two dimensional Lie
algebra spanned by two operators $e_1,e_2$ with the unique
non-trivial commutator $[e_1,e_2]=e_2$ as $L_2$. The algebra $L_4$
is a decomposable Lie algebra and can be written as a semi-direct
sum
\begin{equation} L_4=L_2 \bigoplus e_3 \bigoplus e_4,~~
L_2=<~e_1,e_2~>,~~[e_1,e_2]=e_2.
\end{equation}

{\bf Case $r\ne 0$}. In the case $r \ne 0 $ the generators take
the form
\begin{eqnarray} \label{esr}
e_1=(r-1)U_1+U_3=(r-1)S\frac{\partial}{\partial S}+(r-1)u
\frac{\partial}{\partial u} + \frac{\partial}{\partial t},~~~~
e_2=U_2=e^{r t}\frac{\partial}{\partial u},\nonumber\\
e_3=r U_1+U_3=rS\frac{\partial}{\partial S}+ r u
\frac{\partial}{\partial u}+ \frac{\partial}{\partial t},~~~~~
e_4=U_4=S \frac{\partial}{\partial u}.~~~~~
\end{eqnarray}

{\bf Case $r = 0$}. Using the previous notations we can represent
$L_4$ in the case $r=0$ in the form
\begin{eqnarray} \label{esr0}
e_1=-U_1=-S\frac{\partial}{\partial S} -u \frac{\partial}{\partial
u},~~~~~~e_2=U_2=\frac{\partial}{\partial u},\nonumber\\
e_3=U_3= \frac{\partial}{\partial t},~~~~~~~ e_4= U_4=S
\frac{\partial}{\partial u}.
\end{eqnarray}

Patera and Winternitzs  \cite{PateraWinternitzs1977} looked for
classifications of the sub-algebras into equivalence classes under
their group of inner automorphisms. They also used the idea of
normalization which guarantees that the constructed optimal system
of subalgebras is unique up to the isomorphisms.

This classification allows us to divide the invariant solutions
into non-intersecting equivalence classes. In this way it is
possible to find the complete set of essential different invariant
solutions to the equation under consideration. We use this
classification and give a list of all non-conjugate one-, two- and
three-dimensional subalgebras. The optimal normalized system of
subalgebras to the algebra $L_4$ is listed in Table \ref{optsev}.

\begin{table}
\tbl{\cite{PateraWinternitzs1977} The optimal system of
subalgebras $h_i$ of the algebra $L_4$ where $a\in {\mathbb
R},~~\epsilon=\pm1,~~\phi \in [0,\pi ] $.} {\begin{tabular}{|c|c|}
\hline
{Dimension}&{Subalgebras}\\
\hline
$1$&$h_1=<~e_2~>,~~h_2=<~e_3 \cos {(\phi)}+e_4\sin{(\phi)}~>,$\\
&$h_3=<~e_1+ a(e_3 \cos {(\phi)}+e_4\sin{(\phi)})~>,$\\
&$h_4=<~e_2 +\epsilon(e_3 \cos{(\phi)}+e_4\sin{(\phi)})~>$\\
\colrule
$2$&$h_5=<~e_1+a(e_3 \cos {(\phi)}+e_4\sin{(\phi)}),e_2>,~~h_6=<e_3,e_4~>,$\\
&$h_7=<~e_1 +a(e_3 \cos {(\phi)}+e_4\sin{(\phi)}),e_3 \sin{(\phi)}-e_4\cos{(\phi)}~>,$\\
&$h_8=<~e_2 +\epsilon(e_3 \cos{(\phi)}+e_4\sin{(\phi)}),e_3 \sin{(\phi)}-e_4\cos{(\phi)}~>,$\\
&$h_9=<~e_2,e_3 \sin {(\phi)}-e_4\cos{(\phi)}~>$\\
\colrule
$3$&$h_{10}=<~e_1,e_3,e_4>,~~h_{11}=<~e_2,e_3,e_4~>,$\\
&$h_{12}=<~e_1 +a(e_3 \cos {(\phi)}+e_4\sin{(\phi)}),e_3 \sin {(\phi)}-e_4\cos{(\phi)},e_2~>$\\
\hline
\end{tabular}}
\label{optsev}
\end{table}

 In Table \ref{optsev} we use the operators
$e_1,e_2,e_3$ given by (\ref{esr0}) if $r=0$ and by (\ref{esr}) if
$r \ne 0$.

In correspondence with the set of subalgebras listed in Table
\ref{optsev}, we obtain the complete set of invariant functions
and reduce equation (\ref{sevcov}) to different ODEs using these
functions as dependent and independent variables.

\section{Group-invariant reductions provided by the one-dimensional
symmetry subgroups in the case $r\ne 0 $ }

In this section we study the symmetry reductions of the RAPM model
(\ref{sevcov}) which we obtain using one of the one-dimensional
symmetry subgroups $H_i, i=1,...,4.$ These symmetry subgroups $H_i
\subset G_4$  are generated by the corresponding subalgebras
$h_i,i=1,...,4$ listed in Table \ref{optsev} by a usual
exponential map. We skip the study of invariant reductions to the
two and three dimensional subgroups listed in Table \ref{optsev}
because they only give trivial results for the RAPM model.\\\\
\vspace{5pt}

{\bf Case $H_1$.} This one-dimensional subgroup $H_1$ is generated
by the subalgebra $$ h_1=<~e_2~>=<~e^{r t}\frac{\partial}{\partial
u}~> .$$ It describes a gauge (or evolutionary) symmetry of the
equation. It means that to each solution to equation
(\ref{sevcov}) we can add a term $\alpha e^{r t}$, where $\alpha$
is arbitrary constant. The new function $u(t,S) \to u(t,S)+ \alpha
e^{r t}$ is then still a solution to the equation. This symmetry
does not
give rise to any invariant reductions of equation (\ref{sevcov}).\\
\vspace{5pt}

{\bf Case $H_2$.} We look for the invariants of the subalgebra
$h_2=<~e_3 \cos {(\phi)}+e_4\sin{(\phi)}~>$. In the variables
$(t,S,u)$ we obtain that $h_2$ has the form
\begin{equation}h_2=<~\cos {(\phi)}\frac{\partial}{\partial t}+r \cos {(\phi)}
S\frac{\partial}{\partial S}+ (\cos {(\phi)}~ r ~u +\sin{(\phi)}S
)\frac{\partial}{\partial u}~> .
\end{equation}
The invariants  $z, w $ of the corresponding subgroup $H_2 \subset G_4$ can
be chosen in the form
\begin{equation}
z= S e^{-rt}, ~~w=\frac{u}{S} -\frac{\tau}{r}\ln{S},~~ r\ne
0,~~\tau= \tan{{(\phi)}}, ~{\phi}\in [0, \pi],~{\phi}\ne \pi/2.
\end{equation}
We take the invariants  $z,w$ as the new independent and dependent
variables, respectively, then the PDE (\ref{sevcov}) is reduced to
the ordinary differential equation of the following form
\begin{eqnarray}\label{reduh2}
\left(\tau +r z(z w_{zz}+2w_z) \right) \left(1- \mu
r^{-\frac{1}{3}} \left(\tau + r z(z w_{zz}+2 w_z)
\right)^{\frac{1}{3}} \right)
 +\frac{2 r \tau}{\sigma^2} =0,\\ r\ne 0,~~~~\tau=\tan{{(\phi)}},~~{\phi}\in [0, \pi],~{\phi}\ne \pi/2. \nonumber
\end{eqnarray}
This second order differential equation can be reduced to a first
order equation by the substitution $w_z(z)=v(z)$ which has the
form
\begin{eqnarray}
\left(\tau +r (z^2 v)_{z} \right) \left(1- \mu r^{-\frac{1}{3}}
\left(\tau + r (z^2 v)_{z} \right)^{\frac{1}{3}} \right)
 +\frac{2 r \tau}{\sigma^2} =0.
\end{eqnarray}
From this equation it follows that the expression $(z^2 v)_{z}$ is
a constant. If we denote $(\tau +r (z^2 v)_{z})^{1/3}=p(z)$,  then
for the value $p(z)$ we obtain an algebraic equation of the fourth
order
\begin{equation}\label{poly4}
p^3 \left(1- \mu r^{-\frac{1}{3}} p \right) +\frac{2 r
\tau}{\sigma^2} =0.
\end{equation}
This equation has four roots $q_i, i=1, \dots, 4$. In dependence
on the values of the constants $\mu$ and $\tau$ some of these
roots are real. We denote the real roots by $k_i$. To find
solutions to the ODE (\ref{reduh2}) we have just to integrate two
simple first order differential equations
\begin{equation}
\tau + r (z^2 v)_{z}=k_i^3, ~~w_z(z)=v(z).
\end{equation}
Then to each root $k_i$ the corresponding solutions  to equation
(\ref{reduh2}) are given as two parametric families of functions
\begin{equation}
u(S,t)=\frac{k_i^3}{r} S \ln S - (k_i^3 -\tau) t S + c_1 S + c_2
e^{r t },
\end{equation}
where $ c_1,~c_2 \in {\mathbb R}, r\ne 0, \tau= \tan{{(\phi)}},
{{\phi}} \in
  [0,\pi],~~{\phi}\ne \pi/2.$\\ \vspace{5pt}

{\bf Case $H_3$.} The subalgebra $h_3$ is spanned by the generator
$e_1+ a(e_3 \cos {(\phi)}+e_4\sin{(\phi)})$. In the variables
$(t,S,u)$ it means that we have to do with the subalgebra of the
form
\begin{eqnarray} 
h_3=<~(1+ a \cos {(\phi)})\frac{\partial}{\partial t}+((r-1)+ a r
\cos {(\phi)} ) S \frac{\partial}{\partial S}+ \\\nonumber ((r-1)u
+ a (\cos {(\phi)} r u +\sin{(\phi)}S )\frac{\partial}{\partial
u}~>.
\end{eqnarray}
 The two first invariants of the corresponding subgroup $H_3$ are given by $z,w$
which are connected to variables $(t,S,u)$ by
\begin{equation}
 z= S e^{-(r +\gamma)t}, ~~u(S,t)=S
w(z)+\zeta S \log{S},
\end{equation}
where the constants are $ \gamma=(1+a \cos (\phi))^{-1}$, $\zeta
=\frac{a \sin (\phi)}{r (1 + a \cos (\phi))-1},
a \in {\mathbb R}$, ${\phi}\in [0, \pi] $.\\
Using these expressions we reduce the RAPM equation to an ordinary
differential equation of the form
\begin{equation}
\frac{\sigma^2}{2} \left(z(zw)_{zz}+\zeta \right)\left(1- \mu
\left(z(zw)_{zz}+\zeta\right)^{\frac{1}{3}}\right)
  +r \zeta -\gamma z w_z =0.
\end{equation}
The solutions to this equation can be given in the parametric form
\begin{eqnarray} \label{h3paramne0}
z(\theta)&=& \exp{ \left(  \int \frac{{\rm d} \theta}{k_i(\theta)^3 -\theta -\zeta} \right)},\nonumber\\
 w(\theta)&=&   \int {\frac{\theta {\rm d}
\theta}{k_i(\theta)^3 -\theta -\zeta} },
\end{eqnarray}
where $\theta \in \mathbb R$ is a parameter and $q_i(\theta)$ is
one of the real roots of the fourth order algebraic  equation
\begin{equation} \label{polyh3ne0}
\frac{\sigma^2}{2}k_i(\theta)^3(1-\mu k_i(\theta)) +r \zeta -
\gamma \theta=0.
\end{equation}
\\ \vspace{5pt}

{\bf Case $H_4$.} The subalgebra $h_4$ is spanned by the generator
$e_2+ a(e_3 \cos {(\phi)}+e_4\sin{(\phi)})$. In terms of the
variables $(t,S,u)$ it means that we are dealing with the
subalgebra of the form
\begin{equation} 
h_4=<~ \epsilon \cos {(\phi)}\frac{\partial}{\partial t}+ \epsilon
r \cos {(\phi)}  S \frac{\partial}{\partial S}+ (e^{r t} +
\epsilon (\cos {(\phi)} r u +\sin{(\phi)}S
)\frac{\partial}{\partial u}~>.
\end{equation}
The invariants of the corresponding subgroup $H_4$ are $z$ and
$w$, where
\begin{equation}z= S e^{- rt},~~
u(S,t)=S w(z)+ \left( \frac{\tau}{r} + \frac{\epsilon}{r \cos
(\phi)} z^{-1}\right) S \log{S},
\end{equation}
with $ \tau= \tan (\phi),$ ${\phi}\in [0, \pi],~{\phi}\ne \pi/2$
and $\epsilon=\pm 1$. We take these invariants as new invariant
variables and reduce equation (\ref{sevcov}) to an ODE of the
following form
\begin{eqnarray} \label{reduch4ne0}
&&\frac{\sigma^2}{2} \left(z(z w)_{zz}+ \frac{\tau}{r} +
\frac{\epsilon}{r z \cos (\phi )}\right)\left(1- \mu \left(z(z
w)_{zz} + \frac{\tau}{r} + \frac{\epsilon}{r z \cos (\phi)
}\right)^{\frac{1}{3}}\right) \nonumber\\
&& + \tau + \frac{\epsilon}{z \cos (\phi )} =0.
\end{eqnarray}
If we denote $ p(z)=\left( z(z w)_{zz} + \frac{\tau}{r} +
\frac{\epsilon}{r z \cos (\phi) }\right)^{\frac{1}{3}}$ then for
the value $p(z)$ we obtain an algebraic equation of the fourth
order
\begin{equation} \label{polyh4ne0}
p^3(z) \left(1 - \mu p(z) \right) + \frac{2\tau}{\sigma^2} +
\frac{2 \epsilon }{z \sigma^2 \cos (\phi )} =0.
\end{equation}
This equation has four roots which we denote $q_i, i=1, \dots,4$
as in the case $H_2$.

{\bf Remark.} The roots $q_i$ in this equation differ from the
roots of equation (\ref{polyh3ne0}) or (\ref{poly4}). Still, we
denote here (and later) all real roots of a fourth order algebraic
equation by $k_i$ to show the similar structure of
solutions.\\

Then to each root $k_i(z)$ the corresponding solutions to equation
(\ref{sevcov}) are given as two-parametric families of functions
\begin{eqnarray}
u(S,t)&=&e^{r t} \int{\left(\int{\frac{k_i(z)^3}{z} {\rm
d}z}\right)
 {\rm d}z}+ S\left(\tau t  + c_1 \right)
 \nonumber \\
 &+&
 e^{r t}\left(\frac{\epsilon}{\cos{(\phi)}} t  +c_2 \right),
\end{eqnarray}
where $\tau = \tan(\phi)$, $z= S e^{- rt}$, ${\phi}\in [0,
\pi],~{\phi}\ne \pi/2$, $c_1,c_2 \in {\mathbb R}$
and $\epsilon=\pm 1$.\\
\vspace{5pt}

{\bf The special case of invariant solutions}.\\ \vspace{5pt}

In some cases it is more rewarding  not to take one of the
classical representatives listed in Table \ref{optsev} of the
non-conjugated subalgebras but rather turn to an equivalent one
which gives us a simpler ODE. Let us take a one-dimensional
subalgebra of the form $h=<~e_1+\alpha e_2~>$, where $e_1,e_2$ are
defined by (\ref{esr}). The invariants of the corresponding
subgroup $H$ are defined by the infinitesimal generator
\begin{equation}
 U=e_1+\alpha e_2=(r-1)U_1+U_3+ \alpha U_2, \label{invsyst}
\end{equation}
and can be chosen in the form
\begin{equation}
z=S e^{-(r-1)t}, ~~w=u(S,t) e^{-(r-1)t}- \alpha e^{t}.
\label{invariantsr}
\end{equation}

{\bf Remark}. In the case $r=1$ the dependence of the invariants
$z,w$ on $t$ will be trivial. It means then that $z=S$ is an
invariant and $w=u + \alpha e^t$. On the other hand, the value
$r=1$ implies that on the market $100$ per cent interest rates are
accepted. This is certainly a case which can not be modeled with
the RAPM model. We can, therefore, exclude
the case $r=1$.\\

 We use these invariant functions $z$ and
$w$  to reduce the original equation (\ref{sevcov}) to the ODE of
the form
\begin{equation}
-w +z w_z +\frac{1}{2}\sigma^2 z^2 w_{zz}\left(1- \mu (z
w_{zz}\right)^{1/3})=0. \label{reduc2}
\end{equation}
It is easy to see that this equation does not depend on the
arbitrary parameter $\alpha$ which is included in (\ref{invsyst}).
The second order ODE (\ref{reduc2}) can be reduced to a first
order one
\begin{equation}
v_z-\mu v_z^{4/3} =- \frac{2 v}{\sigma^2 z} \label{reduc1}
\end{equation}
by the substitution
\begin{equation} \label{substr2}
v(z,w)=z w_z-w.
\end{equation}
Equation (\ref{reduc1}) has a parametric solution. We obtain this
solution in the following way. We rewrite equation (\ref{reduc1})
in the form
\begin{equation}
v(z)=- \frac{\sigma^2 }{2} z\left( v_z-\mu v_z^{4/3}\right)
\label{reduc11}=G(z,v_z),
\end{equation}
then the parametric solution to this equation is given by the
solution to the system of equations
\begin{equation}
v(\theta)=G(z(\theta),\theta),~~ z_\theta=
\frac{G_\theta(z,\theta)}{\theta-G_z(z,\theta)}=
-\frac{\sigma^2}{2} \frac{ z \left(1 -  \frac{4}{3} \mu
\theta^{\frac{1}{3}} \right)}{\theta \left(1-\frac{\sigma^2 }{2}
\left(1- \mu \theta^{\frac{1}{3}} \right)\right)} \label{reduc12},
\end{equation}
where $\theta \in \mathbb R$ is a parameter. The system
(\ref{reduc12}) and correspondingly equation (\ref{reduc1}) have
the following solution
\begin{equation} \label{paramreduc1}
v(\theta)=-\frac{\sigma^2}{2} z(\theta) (\theta - \mu
\theta^{4/3}), ~~ z(\theta)= c_1 \left( 1 -\frac{\sigma^2}{2}
\left( 1 - \mu \theta^\frac{1}{3}\right)\right)^{1+ 3 \gamma}
\theta^{- \frac{\sigma^2}{2} \gamma },
\end{equation}
where $\gamma =\left(1-\frac{\sigma^2}{2}\right)^{-1}$ and
$c_1=const$. Using the parametric solution (\ref{paramreduc1}) to
(\ref{reduc1}) we obtain the parametric solution to
(\ref{reduc2}). We used the substitution (\ref{substr2}) which now
takes the form
\begin{equation} \label{substr2t}
v(\theta)=z(\theta) w_z-w=\left( \ln
{z(\theta)}\right)_\theta^{-1} w_\theta - w.
\end{equation}
This is a linear first order differential equation for the
function $w(t)$ and together with the parametric representation of
 $z(\theta)$ (\ref{paramreduc1}) the solution to this equation
gives us the parametric solution to (\ref{reduc2})
\begin{equation}\label{solwsp}
w(\theta)=z(\theta)\left(c_2  + g(\theta)\right), ~~c_2={\rm
const},
\end{equation}
where the function $g(\theta)$ is given by
\begin{eqnarray}\nonumber
g(\theta)=\frac{\sigma^2}{2 \mu^2} \theta^{\frac{1}{3}}\left( -4 +
\frac{\sigma^2}{2} \left(5+ 2\mu \theta^{\frac{1}{3}}\right)
-\frac{\sigma^4}{4} \left( 1+ \frac{\mu}{2} \theta^{\frac{1}{3}}
+\frac{4}{3}\mu^2 \theta^{\frac{2}{3}}\right) \right.
\\
\left. -\frac{\mu^2 \sigma^6}{8} \theta^{\frac{2}{3}} \left( 1-
\mu \theta^{\frac{1}{3}}\right) \right)
+\left(\frac{\mu\sigma^2}{2}\right)^{-3} \left( 4
-\frac{\sigma^2}{2} \right) \left( 1-\frac{\sigma^2}{2}\right)^2
\ln \left( 1- \frac{\sigma^2}{2}\left( 1- \mu \theta^{\frac{1}{3}}
\right)\right) .\nonumber
\end{eqnarray}
Expressions (\ref{solwsp}) and (\ref{paramreduc1}) give a
parametric representation of a solution $w(z)$ to equation
(\ref{reduc2}).

\section{Group-invariant reductions provided by one-dimensional
symmetry subgroups in the case $r = 0 $ }

We repeat the procedure of constructing the invariant solutions to
the RAPM model in the case $r=0$. The general structure of the
optimal system of sub-algebras is the same in both cases but the
form of infinitesimal generators differ. The invariants and the
reductions therefore take another forms.\\ \vspace{5pt}

{\bf Case $H_1^0$. } The generator of the subalgebra $h_1^0$ has a
very simple form $e_2=\frac{\partial}{\partial u}$ in the case
$r=0$. This means that we are dealing  with a subgroup of
translations in the $u$-direction. Hence, to each solution to
equation (\ref{sevcov}) with $r=0$, we can add an arbitrary
constant without destroying the property of the function to be a
solution.
This subgroup does not provide any reduction.\\
\vspace{5pt}

{\bf Case $H_2^0$. } The subalgebra $h_2^0$ has the form
$h_2^0=<~e_3 \cos {(\phi)}+e_4\sin{(\phi)}~>$, it means that in
terms of the variables $(t,S,u)$ we have the subalgebra of the
following type
\begin{equation} h_2^0=<~\cos {(\phi)}\frac{\partial}{\partial t}
+\sin{(\phi)}S \frac{\partial}{\partial u}~> .
\end{equation}
The invariants of the subgroup $H_2^0$ are given by
\begin{equation}
z=S, ~~w=u(S,t) - \tau t S, ~~~\tau=\tan{(\phi)},~~{\phi}\in [0,
\pi],~{\phi}\ne \pi/2.
\end{equation}
If we use the variables $z,w$ as new independent and dependent
variables we obtain the following reduction of the RAPM model
(\ref{sevcov}) with $r=0$
\begin{equation} \label{reduc20}
\frac{\sigma^2}{2} z w_{zz}\left(1- \mu
\left(z~w_{zz}\right)^{\frac{1}{3}}\right)
  + \tau  =0,~~\tau=\tan{(\phi)},~~{\phi}\in [0, \pi],~{\phi}\ne \pi/2.
\end{equation}
We denote $\left(z~w_{zz}\right)^{\frac{1}{3}}=p(z)$ and obtain
for the value $p(z)$ an algebraic fourth order equation
\begin{equation}\label{poly40}
p^3 \left(1- \mu  p \right) +\frac{2 \tau}{\sigma^2} =0.
\end{equation}
As before we denote the real roots of this equation by $k_i$. To
find solutions to the ODE (\ref{reduc20}) we have just to
integrate twice
\begin{equation}
z~w_{zz}=k_i^3.
\end{equation}
Then the corresponding solutions to equation (\ref{reduc20}) are
given by
\begin{equation}
u(S,t)=k_i^3 S \left( \ln S -1 \right) + ~\tau t S + c_1 S + c_2,
\end{equation}
where $\tau=tan{(\phi)}$, $ c_1,~c_2 \in {\mathbb R}, {\phi} \in [0,\pi],{\phi}\ne \pi/2.$\\
\vspace{5pt}

{\bf Case $H_3^0$}. The subalgebra $h_3^0$ for $r=0$ has the form
\begin{equation} h_3^0=<~a \cos {(\phi)}\frac{\partial}{\partial t}-S \frac{\partial}{\partial S}
+(a \sin{(\phi)}S -u )\frac{\partial}{\partial u}~> ,
\end{equation}
where $a \in R$, ${\phi} \in [0,\pi]$ are parameters. The
invariants $z,w$ of the group $H_3^0$ are given by the expressions
\begin{equation}
z= S e^{\delta t},~~ u(S,t)=S w(z) + \zeta S \log{S}, \nonumber
\end{equation}
where the parameters are defined as
\begin{equation}
\delta =(a \cos (\phi))^{-1},~~\zeta= a \sin {(\phi)}, ~ a \in R,~
a \ne 0,~\phi \in [0, \pi],~{\phi}\ne \pi/2, \label{constred03}
\end{equation}
and the reduced equation takes the form
\begin{equation}
\frac{\sigma^2}{2} \left(z(zw)_{zz}+\zeta \right) \left(1- \mu
\left(z(zw)_{zz}+\zeta\right)^{\frac{1}{3}}\right) +\delta z w_z
=0.
\end{equation}
The solutions to this equation can be represented in the
parametric form (\ref{h3paramne0}), where $k_i(v)$ is one of the
real roots of the equation
\begin{equation}
\frac{\sigma^2}{2}k_i(v)^3(1-\mu k_i(v)) + \delta v=0,
\end{equation}
and the parameter $\delta$ is defined in (\ref{constred03}).\\
\vspace{5pt}

{\bf Case $H_4^0$}. The subalgebra $h_4^0$ for $r=0$ has the form
\begin{equation} h_4^0=<~ \epsilon \cos {(\phi)}\frac{\partial}{\partial t}+
(1+\epsilon \sin{(\phi)}S)\frac{\partial}{\partial u}~> ,
\end{equation}
where $\epsilon= \pm 1$, ${\phi} \in [0,\pi]$ are parameters.

The invariants $z,w$ of this subgroup $H_4^0$ are given by the
expressions
\begin{equation}
z= S,~~~~  w(z)=u(S,t) -\tau t S -\frac{\epsilon ~t
}{\cos{(\phi)}},~~\tau=\tan{(\phi)},
\end{equation}
and the RAPM model is reduced to the ODE of the form
\begin{equation}\label{reduch40}
\frac{\sigma^2}{2} z^2 w_{zz} \left(1 - \mu \left( z
w_{zz}\right)^{\frac{1}{3}} \right) +\tau z
+\frac{\epsilon}{\cos{(\phi)}}=0,
\end{equation}
where $\tau=\tan{(\phi)}$, ${\phi}\in [0, \pi],~{\phi}\ne \pi/2,
~\epsilon= \pm 1$. The structure of equation (\ref{reduch40}) is
very similar to previous cases and we can use similar tools to
solve it. We first substitute $(z w_{zz})^{1/3}=p(z)$. Then for
the function $p(z)$ we obtain a fourth order algebraic equation
but now its coefficients depend on the variable $z$
\begin{equation}\label{polyh440}
p(z)^3 \left(1- \mu  p(z) \right) +\frac{2 \tau}{\sigma^2}
+\frac{2 \epsilon}{z \sigma^2\cos{(\phi)}} =0,
\end{equation}
where $\tau=\tan{(\phi)}$, ${\phi}\in [0, \pi],~{\phi}\ne \pi/2,~
\epsilon= \pm 1$. For each real root $k_i(z)$ of this equation we
have then to solve a linear ODE
\begin{equation}
z~w_{zz}=k_i(z)^3.
\end{equation}
The corresponding invariant solutions to (\ref{sevcov}) then have
the form
\begin{equation}
u(t,S)= \int{\left(\int{\frac{k_i(S)^3}{S} {\rm d}S}\right){\rm
d}S} +\tan{(\phi)}~ t ~S +\frac{\epsilon}{\cos{(\phi)}} +c_1 S
+c_2,
\end{equation}
where $c_1,~c_2 \in {\mathbb R}, ~{\phi} \in [0,\pi],~{\phi}\ne \pi/2,~ \epsilon= \pm 1.$\\
The expressions for these solutions are rather lengthy and because
of which they are omitted here.

\section{Conclusion}

In the previous sections we found the complete series of
reductions of the RAPM model. In this way the partial differential
equation (\ref{sevcov}) is reduced to ordinary differential
equations. Using the optimal system of subalgebras (Table
\ref{optsev}) allowed us to present the complete set of the
non-equivalent reductions of equation (\ref{sevcov}) up to the
transformations of the group $G_4$. In all cases it is possible to
solve these ODEs and to obtain the exact or parametric
representations of solutions to the RAPM model. We deal with the
very seldom case that we can compare structures of non-equivalent
invariant solutions since they are given in exact or parametric
forms. Each of these solutions contains two integration parameters
and some free parameters connected with the corresponding
subgroup. This reasonable set of parameters allowes one to
approximate a wide class of boundary conditions.\\
The RAPM model (\ref{sevcov}) possesses a non-trivial analytical
and singular-perturbed algebraic structure. There exist rather few
methods to study equations of such high complexity.
 An application of
both analytical and numerical methods to singular-perturbed
equations is a highly non-trivial task. The RAPM model was studied
before in detail with numerical methods in
\cite{JandarckaSevcovic} and in \cite{Sevcovic}. The authors of
\cite{JandarckaSevcovic} derive a robust numerical scheme for
solving equation (\ref{sevcov}) and perform extensive numerical
testing of the model and compare the results to real market data.
In \cite{Sevcovic} {{\v{S}}ev{\v{c}}ovi{\v{ c}}} studies the free
boundary problem for the RAPM model and provides a description of
the early exercise boundary for American style Call options with
floating strike. He proposed a numerical method based on the
finite difference approximation combined with an operator
splitting technique for numerical approximation of the solution
and computation of the free boundary condition position.\\
On the other hand the Lie group analysis of the RAPM model which
we provide in this paper gives us  a more general, alternative
point of view on the structure of this equation. It opens  the
possibility to exploit the Lie algebraic structure of the equation
and may be helpful to improve another methods.

\section*{Acknowledgements}
The author is grateful to Michael Nechaev and Alexandr Yanovski
for helpful discussions.

\end{document}